\documentclass[useAMS,usenatbib]{mnras}
\usepackage{graphicx}
\usepackage{multirow, bigstrut}%
\usepackage{epstopdf}%
\usepackage{color}%
\usepackage{soul}%
\usepackage{amssymb}
\usepackage{amsmath}
\usepackage[usenames,dvipsnames]{xcolor}%

\def\spose#1{\hbox to 0pt{#1\hss}}
\def\gtsimm{\mathrel{\spose{\lower 3pt\hbox{$\sim$}}
        \raise 2.0pt\hbox{$>$}}}

\begin{document}
\title[$\chi=10^{3}$ shock/wind-cloud comparison]{A comparison of shock-cloud and wind-cloud interactions: Effect of increased cloud density contrast on cloud evolution}
\author[K. J. A. Goldsmith and J. M. Pittard] 
  {K. J. A. ~Goldsmith \thanks{pykjag@leeds.ac.uk} and J. M. ~Pittard\\
   School of Physics and Astronomy, University of Leeds, 
   Woodhouse Lane, Leeds LS2 9JT, UK}
   
\date{Accepted ... Received ...; in original form ...}

\pagerange{\pageref{firstpage}--\pageref{lastpage}} \pubyear{2016}
\newtheorem{theorem}{Theorem}[section]
\label{firstpage}

\maketitle

\begin{abstract}
The similarities, or otherwise, of a shock or wind interacting with a cloud of density contrast $\chi=10$ were explored in a previous paper. Here, we investigate such interactions with clouds of higher density contrast. We compare the adiabatic hydrodynamic interaction of a Mach 10 shock with a spherical cloud of $\chi=10^{3}$ with that of a cloud embedded in a wind with identical parameters to the post-shock flow. We find that initially there are only minor morphological differences between the shock-cloud and wind-cloud interactions, compared to when $\chi=10$. However, once the transmitted shock exits the cloud, the development of a turbulent wake and fragmentation of the cloud differs between the two simulations. On increasing the wind Mach number we note the development of a thin, smooth tail of cloud material, which is then disrupted by the fragmentation of the cloud core and subsequent ``mass-loading'' of the flow. We find that the normalised cloud mixing time ($t_{mix}$) is shorter at higher $\chi$. However, a strong Mach number dependence on $t_{mix}$ and the normalised cloud drag time, $t_{drag}'$, is not observed. Mach-number-dependent values of $t_{mix}$ and $t_{drag}'$ from comparable shock-cloud interactions converge towards the Mach-number-independent time-scales of the wind-cloud simulations. We find that high $\chi$ clouds can be accelerated up to $80-90$\% of the wind velocity and travel large distances before being significantly mixed. However, complete mixing is not achieved in our simulations and at late times the flow remains perturbed.
\end{abstract}

\section{Introduction} 
\label{intro}
The interstellar medium (ISM) is a dynamic entity, the study of which can allow insights into the nature of the ISM itself \citep[see e.g.][]{Elmegreen04, MacLow04, Scalo04, McKee07, Hennebelle12, Padoan14}, as well as processes such as the formation of filamentary structures that are prevalent throughout the ISM. The interaction of hot, high-velocity, tenuous flows (e.g. shocks and winds) with much cooler, dense clumps of material (i.e. clouds), shapes and evolves these clouds and, ultimately, destroys them. A review of shock-cloud studies is presented in \citet{Pittard16b}, whilst an equivalent review of wind-cloud studies can be found in \citet{Goldsmith17}.

Under certain circumstances, flows interacting with clouds can lead to the formation of tail-like morphologies or filamentary structures. Observations have shown these to occur from the small scale, such as comet plasma tails \citep[e.g.][]{Brandt00, Buffington08, Yagi15} to much larger scales, e.g. H$\alpha-$emitting filaments occurring within galaxies. Tails have been observed in NGC 7293 in the Helix nebula (\citealt{ODell05, Hora06, Matsuura07, Matsuura09, Meaburn10})  (see also \citet{Dyson06} for a corresponding numerical study) and also in the Orion Molecular Cloud OMC1 \citep{Allen93, Schultz99, Tedds99, Kaifu00, Lee00}. Tail-like structures have also been found in Galactic winds \citep{Cecil01, Ohyama02, Cecil02, Crawford05, McClure12, McClure13, Shafi15}.

Numerical shock/wind-cloud studies which have had either a particular focus on, or have noted, the formation of tails include \citet{Strickland00, Cooper08, Cooper09, Pittard09, Pittard10}; and \citet{Banda16}, whilst \citet{Pittard11} investigated the formation of tails in shell-cloud interactions. \citet{Pittard09, Pittard10}, for example, noted the formation of tail-like structures in 2D shock-cloud interactions where the cloud had a density contrast $\chi=10^{3}$ and a high shock Mach number and suggested that this was because the stripping of material was more effective at higher Mach numbers due to the faster growth of Kelvin-Helmholtz (KH) and Rayleigh-Taylor (RT) instabilities. They found that well-defined tails formed only for density contrasts $\chi \gtrsim 10^{3}$, but developed for a variety of Mach numbers. 

In contrast, whilst there are a large number of wind-cloud simulations in the literature, very few have considered clouds with density contrasts of $10^{3}$ or greater. Those that have \citep[e.g.][]{Murray93, Schiano95, Vieser07, Cooper09, Scannapieco15, Banda16} have tended not to vary the wind Mach number. \citet{Banda16}, for example, noted the realistic nature of higher cloud density contrasts (i.e. $\chi>100$) but limited their adiabatic calculations to winds of Mach number 4. 

In \citet{Goldsmith17} (hereafter denoted as Paper I) we compared shock-cloud and wind-cloud simulations using similar flow parameters for a cloud density contrast $\chi=10$, and explored the effect of increasing the wind Mach number on the evolution of the cloud. In that study, we found there to be significant differences between shock-cloud and wind-cloud interactions in terms of the nature of the shock driven through the cloud and the axial compression of the cloud, and noted that the cloud mixing time normalised to its crushing timescale increased for increasing wind Mach number until it reached a plateau due to Mach scaling. In addition, we also found that clouds in high Mach number winds were capable of surviving for longer and travelling considerable distances. In the current paper, we extend our investigation to clouds with a density contrast higher than that of the first paper ($\chi=10^{3}$) and again compare between simulations where the wind Mach number is varied. We also make comparisons between the current work and Paper I.

The outline of this paper is as follows: in Section 2 we introduce the numerical method and describe the initial conditions, whilst in Section 3 we present our results. Section 4 provides a summary of our results and our conclusions.

\section{The numerical setup}
The calculations in this study were performed on a 2D $RZ$ axisymmetric grid using the \textsc{mg} adaptive mesh refinement hydrodynamical code, where refinement and de-refinement are performed on a cell-by-cell basis (see Paper I for a detailed description of the refinement process). \textsc{mg} solves the Eulerian equations of hydrodynamics, the full set of which can be found in Paper I. The code uses piecewise linear cell interpolation to solve the Riemann problem at each cell interface in order to determine the conserved fluxes for the time update. The scheme is second-order accurate in space and time and uses a linear solver in most instances \citep{Falle91}. 

The effective resolution is quoted as that of the finest grid, $R_{cr}$, where `cr' denotes the number of cells per cloud radius on the finest grid. All simulations were performed at a resolution of $R_{128}$, which has been found to be the minimum necessary for key features in the flow to be adequately resolved and for the morphology and global statistical values to begin to show convergence (e.g. \citealt{Klein94, Niederhaus07, Pittard09, Pittard16b}). As before, we measure all length scales in units of the cloud radius, $r_{c}$, where $r_{c}=1$, whilst velocities are measured in terms of the shock speed through the background medium, $v_{b}$ ($v_{b}=13.6$, in computational units). Measurements of the density are given in terms of the density of the background medium, $\rho_{amb}$. The numerical domain is set to be large enough so that the main features of the interaction occur before cloud material reaches the edge of the grid. Table~\ref{Table1} details the grid extent for each of the simulations. 

We make the following assumptions in order to maintain simplicity: the cloud is adiabatic (with $\gamma = 5/3$) and we ignore the effects of thermal conduction, magnetic fields, self-gravity, and radiative cooling. Our assumption of adiabacity is consistent with the small-cloud-limit, whereby the cloud-crushing time-scale is much shorter than the cooling time-scale (cf. \citealt{Klein94}). Non-radiative interactions between shocks/winds and clouds are expected in the ISM \citep{McKee75}. We further justify our simplified set-up by noting that our primary goal is to provide an initial comparison of shock-cloud and wind-cloud simulations and the similarities/differences between the two types of interaction are better isolated without the introduction of additional processes. We do not, therefore, concern ourselves at this stage with the detail of the processes which led to the cloud being embedded in the wind, nor with the effects of additional processes (e.g. radiative cooling) on the interaction. It should, however, be noted that 3D calculations are necessary in future work and that they are expected to produce slightly different morphologies and statistical values once non-axisymmetric instabilities become important at late times \citep[e.g. $t > 5 \, t_{cc}$][]{Pittard16b}. More realistic 3D comparative studies that include radiative cooling should be considered in the future.

\subsection{The shock-cloud model}
\label{shock}
Our reference simulation is the shock-cloud model $c3shock$ (see Section~\ref{results} for the model naming convention). The simulated cloud is an idealised sphere and is assumed to have sharp edges (see e.g. \citealt{Nakamura06, Pittard16b} for a discussion of how cloud density profiles affect the formation of hydrodynamic instabilities), in contrast to previous shock-cloud studies that used a soft edge to the cloud (e.g. \citealt{Pittard16b}), and is initially in pressure equilibrium with the surrounding stationary ambient medium. The simulations are described by the shock Mach number, $M_{shock}=10$, and the density contrast between the cloud and the stationary ambient medium, $\chi=10^{3}$. The shock-cloud simulation begins with the shock initially located at $z=1$ (the shock propagates in the negative $z$ direction) and the cloud centred on the grid origin $r,\,z=(0,0)$.

The post-shock\footnote{We use the subscript \textit{ps/wind} to denote quantities related to either the post-shock flow or the wind.} density, pressure, and velocity for the shock-cloud case relative to the pre-shock ambient values and to the shock speed are $\rho_{ps/wind}/\rho_{amb} = 3.9$, $P_{ps/wind}/P_{amb} = 124.8$, and $v_{ps/wind}/v_{b} = 0.74$, respectively.

\subsection{The wind-cloud model}
In order to simulate a wind-cloud interaction, we begin by removing the initial shock and fill the domain external to the cloud with the same post-shock flow properties. At the start of the simulation, the cloud is instantly surrounded by a wind of uniform speed and direction, in line with previous wind-cloud studies (e.g. \citealt{Banda16}). Since this is an idealised scenario as a first step towards more realistic simulations, we simplify the initialisation of the wind and make the following assumptions: a) the wind is associated with the post-shock flow properties of the shock-cloud model (i.e. we simulate a mildly supersonic wind using exactly the same post-shock flow conditions as used in the shock-cloud model) and b) that it completely surrounds the cloud at time zero. Our aim is to provide comparable initial conditions for both interactions before any of the wind parameters are changed. This means that the cloud is initially under-pressured compared to the wind. Astrophysically, this implies that the wind switches on rapidly.

Although the initial cloud density is the same in both the shock-cloud and wind-cloud simulations, the density contrast between the cloud and the \emph{wind} in the latter case ($\chi'$) is given by factoring off the value of the post-shock density jump from the value of $\chi$, i.e. $\chi'=\chi/3.9$ (see Section~\ref{shock}).

In addition to the parameters described in Section~\ref{shock}, the wind-cloud simulations are also described by the effective Mach number of the wind, $M_{ps/wind}$, given by
\begin{equation}
M_{ps/wind} = \frac{v_{ps/wind}}{c_{ps/wind}} \, ,
\end{equation}
where $c_{ps/wind} = \sqrt{\gamma \frac{P_{ps/wind}}{\rho_{ps/wind}}}$ is the adiabatic sound speed of the post-shock flow/wind. For our initial wind-cloud simulation (model $c3wind1$), $M_{ps/wind}=1.36$. Since the initial, unshocked cloud pressure is equal to $P_{amb}$, and $P_{amb} \ll P_{ps/wind}$, the cloud does not start off in pressure equilibrium with the wind and is thus under-pressured with respect to the flow. Over the course of one cloud-crushing time-scale the cloud pressure increases until it is equal to or slightly greater than the pressure of the surrounding wind. It should be noted that the wind can travel a long way in the `cloud-crushing time' due to the high density contrast of the cloud. This is a different set-up to other wind-cloud studies (e.g. \citealt{Schiano95}) where the simulations begin with the cloud already in approximate ram pressure equilibrium with the wind, but is necessary in order to allow a more direct comparison to our shock-cloud simulation.

The value of the wind velocity, $v_{ps/wind}$, is given in Section~\ref{shock}. In order to explore the effect of an increasing Mach number on the interaction, the velocity of the flow, $v_{ps/wind}$, is increased by factors of $\sqrt{10}$, $\sqrt{100}$, and $\sqrt{1000}$ in order to increase $M_{ps/wind}$. Values of the wind Mach number are given in Table~\ref{Table1}.

\begin{table}
\caption{The grid extent for each of the simulations presented in this paper (see \S 3 for the model naming convention). $M_{ps/wind}$ denotes the effective Mach number of the post-shock flow/wind. Length is measured in units of the initial cloud radius, $r_{c}$.}
\label{Table1}
\begin{tabular}{@{}lccc}
  \hline
 Simulation & $M_{ps/wind}$ &  $R$ & $Z$   \\
    \hline
c3shock & 1.36 &  $0 < R < 20 $& $-400 < Z < 5 $ \\
c3wind1 & 1.36 &  $0 < R< 30$ & $-700 < Z < 5$ \\
c3wind1a & 4.30 &  $0 < R< 30$ & $-700 < Z < 5$ \\
c3wind1b & 13.6 &  $0 < R< 35$ & $-800 < Z < 5$ \\
c3wind1c & 43.0 & $0 < R < 35$ & $-800 < Z < 5$ \\
  \hline
 \end{tabular}
\end{table}

\subsection{Global quantities}
The evolution of the cloud can be monitored through various integrated quantities \citep[see][]{Klein94, Nakamura06, Pittard09, Pittard16b, Goldsmith17}. These include the core mass of the cloud ($m_{core}$), mean velocity in the $z$ direction ($\langle v_{z, cloud} \rangle$), and cloud centre of mass in the $z$ direction ($\langle z_{cloud} \rangle$). In addition, the morphology of the cloud can be described by the effective radii of the cloud in the radial ($a$) and axial ($c$) directions, defined as
\begin{equation}
a = \left(\frac{5}{2} \langle r^{2}\rangle \right)^{1/2}, \; \; \; c = [5(\langle z^{2}\rangle - \langle z\rangle^{2})]^{1/2} \, ,
\end{equation}
in addition to their ratio.

We use an advected scalar, $\kappa$, to trace the evolution of the cloud in the flow and distinguish between the cloud core and the ambient background. Therefore, we are able to compute each of the global quantities for either the cloud core and associated fragments (using the subscript `core') or the entire cloud plus regions where cloud material is mixed into the surrounding flow (using the subscript `cloud'). Motion is defined with respect to the direction of shock/wind propagation along the $z$ axis, with motion in that direction being termed `axial' and motion perpendicular to that as `radial'.

\subsection{Time-scales}
We use the `cloud-crushing time' given by \citet{Klein94} for the initial shock-cloud simulation:
\begin{equation}
t_{cc} = \frac{\sqrt{\chi} \,r_{c}}{v_{b}}\, .
\end{equation}
For the wind-cloud simulations, this time-scale is redefined according to the post-shock flow/wind velocity:
\begin{equation}
t_{cc} = \frac{C\, \sqrt{\chi}\,r_{c}}{v_{ps/wind}}\, ,
\end{equation}
where the constant $C$ is given by the ratio of the post-shock flow/wind velocity to the velocity of the shock through the unshocked medium, $v_{ps/wind}/v_{b}$. The value of the constant depends on the value of the shock Mach number ($M_{shock}=10$ in this work) used in the shock-cloud simulation, against which the wind simulations are compared. Thus, for our initial shock and wind simulations, models $c3shock$ and $c3wind1$, the value of $C = 0.74$ and is specific to this Mach number and our adopted value of $\gamma$. The value of $C$ is also dependent on the value of $v_{ps/wind}$ which, in our later wind-cloud models, is varied, resulting in differing values of $C$. Therefore, $t_{cc}$ also varies depending on the particular simulation under consideration. Values for the cloud-crushing time-scale for each simulation are given in Table~\ref{Table3}.

Several other time-scales are used, including the ``drag time'', $t_{drag}$; the ``mixing time'', $t_{mix}$, and the cloud ``lifetime'', $t_{life}$ (see Paper I for a more detailed description of these time-scales). In all of the following our timescales are normalised to $t_{cc}$. Time zero in our calculations is defined as the time at which the intercloud shock is level with the leading edge of the cloud in the shock-cloud case. In the wind-cloud case, the simulation begins with the cloud already surrounded by the flow.
    
\section{Results}
\label{results}
In this section we begin by examining the shock-cloud interaction, model $c3shock$, in terms of the morphology of the cloud and then, maintaining the same initial parameters, compare this to our standard wind-cloud interaction, model $c3wind1$. We then consider the interaction when the Mach number of the wind is increased (models $c3wind1a$ to $c3wind1c$).  

At the end of this section we explore the impact of the interaction on various global quantities. In Paper I we used a naming convention such that the higher velocity wind-cloud simulations were described from ``wind1a'' to ``wind1c''. Thus, in order to compare between the two papers we retain a similar naming convention such that $c3shock$ refers to a shock-cloud simulation with $\chi = 10^{3}$. The ``$1a$'' in model $c3wind1a$, for example, indicates that the interaction has an increased wind Mach number compared to model $c3wind1$. 

\subsection{Shock-cloud interaction}
\begin{figure*}
\centering
\begin{tabular}{c}
\includegraphics[width=170mm]{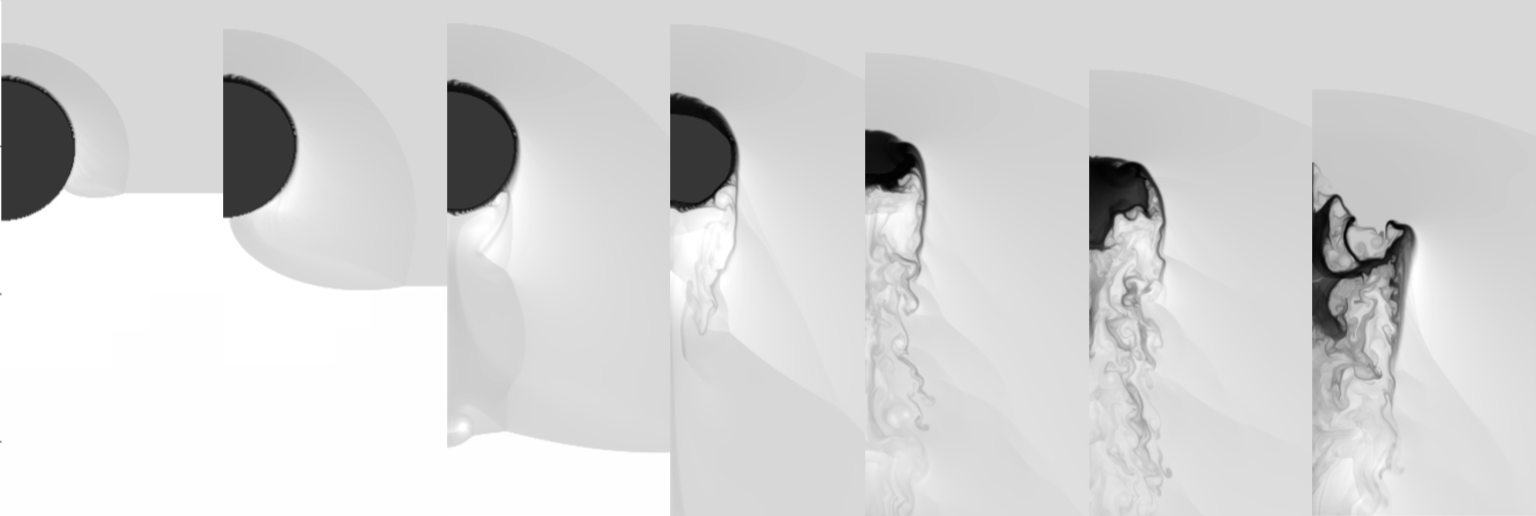}\\
\end{tabular}
\caption{The time evolution of the logarithmic density for model $c3shock$. The greyscale shows the logarithm of the mass density, from white (lowest density) to black (highest density). The density in this and subsequent figures has been scaled with respect to the ambient density, so that a value of 0 represents the value of $\rho_{amb}$ and 1 represents $10 \times \rho_{amb}$. The density scale used for this figure extends from 0 to 3.8. The evolution proceeds left to right with $t=0.043 \,t_{cc}$, $t=0.084 \,t_{cc}$, $t=0.16 \,t_{cc}$, $t=0.31 \,t_{cc}$, $t=1.2\,t_{cc}$, $t=2.0\,t_{cc}$, and $t=3.6 \,t_{cc}$. The $r$ axis (plotted horizontally) extends $3\,r_{c}$ off-axis. All frames show the same region ($-5 < z < 2$, in units of $r_{c}$) so that the motion of the cloud is clear. Note that in this and similar figures the $z$ axis is plotted vertically, with positive towards the top and negative towards the bottom.}
\label{Fig1}
\end{figure*}

Figure~\ref{Fig1} shows plots of the logarithmic density as a function of time for model $c3shock$. The evolution of the cloud broadly proceeds as per model $c1shock$ in Paper I (where $M_{shock}=10$ and $\chi=10$) in that the cloud is initially struck on its leading edge, causing a shock to be transmitted through the cloud whilst the external shock sweeps around the cloud edge, and a bow shock is formed ahead of the leading edge of the cloud. There are a number of differences between the two models, as detailed below. 

The rate at which the transmitted shock progresses through the cloud is considerably slower than the comparable simulation in Paper I; in that paper, the shock was also much flatter whereas model $c3shock$ has a semi-flat shock, the end of which curves around the cloud flank (see fourth panel of Fig.~\ref{Fig1}). The slowness of the transmitted shock and its progress through the cloud in the current simulation is attributed to the increased density of the cloud compared to model $c1shock$. 

Initially, the slow progress of the transmitted shock through the cloud means that the cloud appears to undergo little immediate compression in either the axial or radial directions, in contrast to the cloud in Paper I which was flattened into an oblate spheroid even as the external shock was sweeping around the outside. However, when this is measured in units of $t_{cc}$, maximum compression of the cloud in the axial direction takes place by $t \simeq 1\, t_{cc}$ (cf. panels 4 and 5 of Fig.~\ref{Fig1}).  

The surface of the cloud in the current simulation from the outset is not smooth \citep[compared to the cloud edge in e.g.][]{Pittard09, Pittard10, Pittard16b}. The rapid development of such small instabilities is attributed to the fact that we used a sharp edge to our cloud (see \citet{Pittard16b} for a discussion of how soft cloud edges can hinder the growth of KH instabilities). It is also notable that the cloud moves downstream at a slightly slower rate than would be expected in comparison with previous inviscid shock-cloud calculations (cf. figure 4 in \citet{Pittard09}). This difference is likely to be due to the smooth edge given to the cloud in e.g. \citet{Pittard09} which results in the cloud having slightly less mass than in our model.

The third panel of Fig.~\ref{Fig1} shows that the external shock has reached the $r=0$ axis and cloud material is being ablated from the back of the cloud into the flow. The sheer across the surface of the cloud induces the growth of instabilities, leading to a thin layer of material being drawn away from the side of the cloud and funnelled downstream. At this point, the transmitted shock is still progressing through the cloud. With the transmitted shock curving around the edge of the cloud and also moving in from the rear, the cloud begins to exhibit a shell-like morphology, with a shocked denser outer layer encompassing the unshocked interior. This is a relatively short-lived morphology, since by $t = 1.2\, t_{cc}$ the shocked parts of the cloud collapse into each other, and the transmitted shock has exited the cloud and accelerated downstream. Cloud material is then ablated by the flow and expands supersonically downstream, forming a long and turbulent wake. The cloud core, however, remains relatively intact after the formation of the turbulent wake and persists for some time as a distinct clump (until $t\approx 5.2 \,t_{cc}$, when it starts to become more elongated and drawn-out along the axial direction). This behaviour differs from the $\chi=10$ cloud investigated in Paper I, where the cloud was destroyed much more rapidly. However, it is in better agreement with inviscid simulations presented in \citet{Pittard09}, who showed that clouds with $\chi=10^{3}$ and a shock Mach number of 10 form a turbulent wake, and that the mass loss at later times resembles a a single tail-like structure (see figures 4 and 7 of that paper).

\subsection{Wind-cloud interaction}
\subsubsection{Comparison of wind-cloud and shock-cloud interactions}
\begin{figure*}
\centering
\begin{tabular}{c}
\includegraphics[width=170mm]{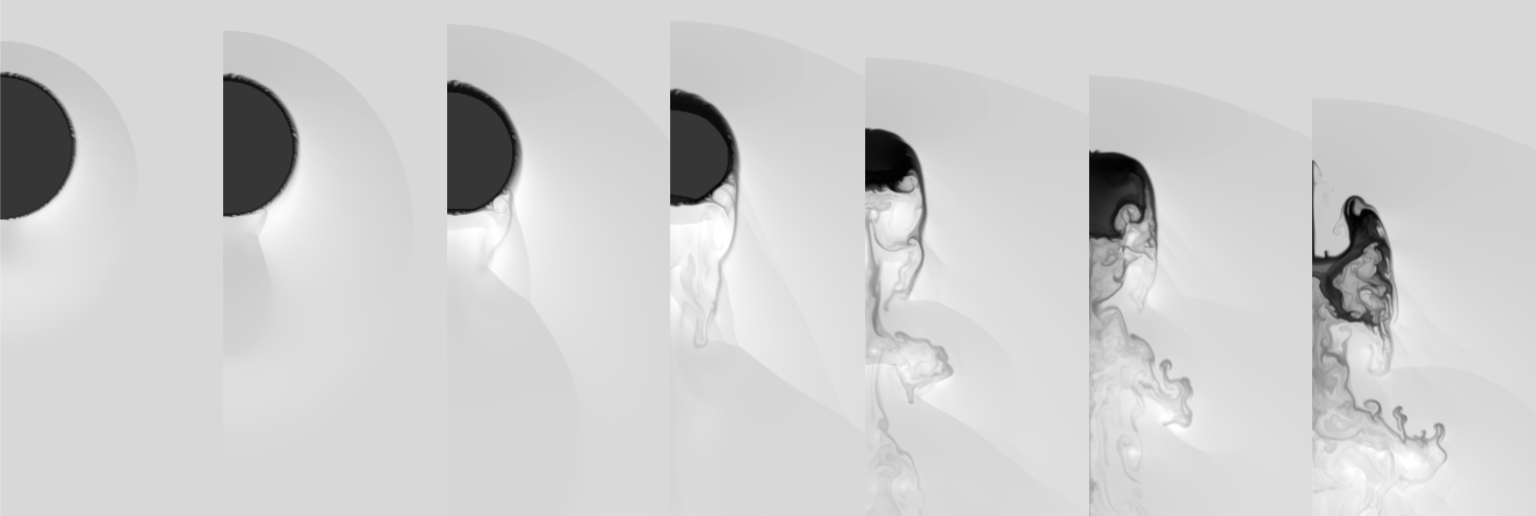}\\
\end{tabular}
\caption{The time evolution of the logarithmic density for model $c3wind1$. The greyscale shows the logarithm of the mass density, scaled with respect to the ambient medium. The density scale used in this figure extends from 0 to 3.8. The evolution proceeds left to right with $t=0.042 \,t_{cc}$, $t=0.077 \,t_{cc}$, $t=0.15 \,t_{cc}$, $t=0.30 \,t_{cc}$, $t=1.2\,t_{cc}$, $t=2.0\,t_{cc}$, and $t=3.6 \,t_{cc}$. All frames show the same region ($-5 < z < 2$, $0 < r < 3$, in units of $r_{c}$) so that the motion of the cloud is clear.}
\label{Fig2}
\end{figure*}

Figure~\ref{Fig2} shows plots of the logarithmic density as a function of time for the wind-cloud case with $M_{wind}=1.36$ ($c3wind1$). Here, the wind density, pressure, and velocity values are exactly the same as the post-shock flow values in model $c3shock$. 

As with models $c1shock$ and $c1wind1$ in Paper I, $c3shock$ and $c3wind1$ show broad similarities (cf. Figs.~\ref{Fig1} and ~\ref{Fig2}). Both clouds have very similar morphologies and there is little to tell them apart, at least initially. However, there are subtle differences between the two models once the initial shock has progressed around the edge of the cloud. For example, the RT instability that develops on the cloud's leading edge behaves differently to that in model $c3shock$. This is due to an area of very low pressure in the shock-cloud case that is situated at the outside (right-hand) edge of the `finger' of cloud material forming due to the RT instability. This low-pressure area is absent in the wind-cloud case. This means that the RT finger is channelled more upstream in the wind-cloud model but expands more radially in the shock-cloud model (see the last 3 panels in Figs.~\ref{Fig1} and~\ref{Fig2}). Furthermore, the flow past the cloud in the wind-cloud case is reasonably uniform, whereas that in the shock-cloud case sweeps around the RT finger and helps to push cloud material outwards in the radial direction. This means that the transverse radius of the cloud grows more quickly in model $c3shock$ compared to $c3wind1$ (see the final panel in Figs.~\ref{Fig1} and~\ref{Fig2}, and also~\ref{Fig4}e). However, in model $c3shock$ the transverse radius of the cloud does not grow any further after $t=3.6\,t_{cc}$, whereas in model $c3wind1$ it continues to do so and by $t=5\,t_{cc}$ it is greater than in model $c3shock$. The continued lateral growth of the cloud in model $c3wind1$ coincides with a greater fragmentation of the core and a more rapid reduction in core mass, so that between $t=5-8\,t_{cc}$ the core mass in $c3wind1$ is less than that in $c3shock$ (see Fig.~\ref{Fig4}a). 

Once the transmitted shock has exited the cloud, the cloud in model $c3wind1$ develops a long, low-density, turbulent wake similar to that in model $c3shock$ (but much less dense) in the downstream direction.\footnote{At late times an axial artifact develops in models $c3shock$ and $c3wind1$. This is visible in the final panels of Figs.~\ref{Fig1} and ~\ref{Fig2} and is seen protruding upstream. Such artifacts are sometimes seen in 2D axisymmetric simulations and occur purely due to the nature of the scheme (fluid can become `stuck' against the boundary). However, it does not appear to influence the rest of the flow and can be safely ignored in our work.} Unlike the cloud in model $c3shock$, the cloud core in model $c3wind1$ is not drawn out along the $z$ direction, and once the core fragments the turbulent wake is disrupted by mass-loading of the core into the flow (not shown).

In comparison to model $c1wind1$ in Paper I, the RT instability in model $c3wind1$ expands upstream as opposed to the radial direction. This effect is caused by shock waves moving through the cloud, once the transmitted shocks from the front and rear of the cloud cross each other. Another difference between our $c3wind1$ simulation and the $c1wind1$ simulation in Paper I is that the rear edge of the cloud is not forced upwards to the same extent due to the action of shocks driven into the back of the cloud (cf. the second panel of Fig.~\ref{Fig2} at $t=0.077\,t_{cc}$ with the second panel of figure 2 in Paper I at $t=0.82\,t_{cc}$). A turbulent wake is not seen in model $c1wind1$ in Paper I.

The evolution of the cloud in model $c3wind1$ bears some similarities to the adiabatic spherical cloud in the wind-cloud study by \citet{Cooper09}, where mass is immediately ablated from the back of the cloud in the form of a long sheet of material and moves downstream in a thin, turbulent tail (see the left-hand panels of figure 7 in \citet{Cooper09} showing the logarithmic density of the cloud, in a $M_{wind}=4.6$ and $\chi = 910$ simulation). Their cloud showed a large expansion in the transverse direction, with cloud material being torn away from the core in all directions and mixed in with the flow, i.e. comparable behaviour to our model $c3wind1$. Such fragmentation of the cloud core is dissimilar to the evolution of the cloud in model $c3shock$.

\subsubsection{Effect of increasing $M_{wind}$ on the evolution}
Compared to model $c3wind1$, models $c3wind1a$, $c3wind1b$, and $c3wind1c$ display a long-lasting and supersonically-expanding cavity located to the rear of the cloud (similar to the higher wind Mach number simulations in Paper I) and a reduced stand-off distance between the cloud and the bow shock; these features are due to the increase in wind velocity and Mach number in these models. 

There is much greater pressure at the leading edge of the cloud in the higher $M_{wind}$ simulations. The density jump at the bow shock in the higher $M_{wind}$ simulations is also greater, and the stand-off distance between the bow shock and the leading edge of the cloud smaller, than in model $c3wind1$. The greater compression at the bow shock reduces the flow velocity (normalised to $v_{ps/wind}$) around the edge of the cloud, leading to a reduction in the growth rate of instabilities and decreased stripping of cloud material from the side of the cloud (when time is normalised to $t_{cc}$). The evolution of the cloud in the higher $M_{wind}$ simulations, therefore, is different to that in model $c3wind1$, especially at low values of the cloud-crushing time-scale. As in Paper I, the higher $M_{wind}$ simulations have very similar morphologies, at least until around $t \approx 1.8\, t_{cc}$. This is due to the presence of the highly-supersonic cavity (as opposed to the area of low pressure behind the cloud in model $c3wind1$) which alters the way the wind flows around the cloud flanks. Instead of being focussed on the $r=0$ axis immediately behind the cloud as in model $c3wind1$, the flow is deflected further downstream away from the cloud edge leading to a much lower pressure jump behind the cloud and restricting secondary shocks from being driven into the rear of the cloud. Thus, there is less turbulent stripping of cloud material from the rear of the cloud in these simulations compared to model $c3wind1$.

\begin{figure*}
\centering
\begin{tabular}{c}
\includegraphics[width=170mm]{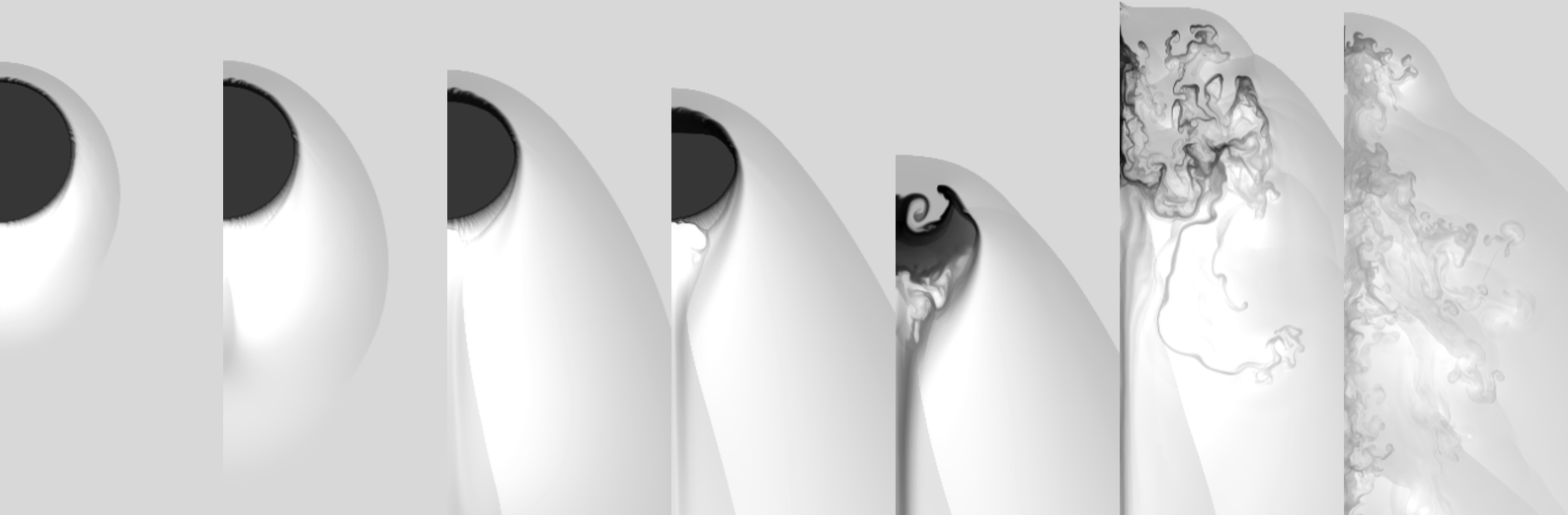} \\
\includegraphics[width=170mm]{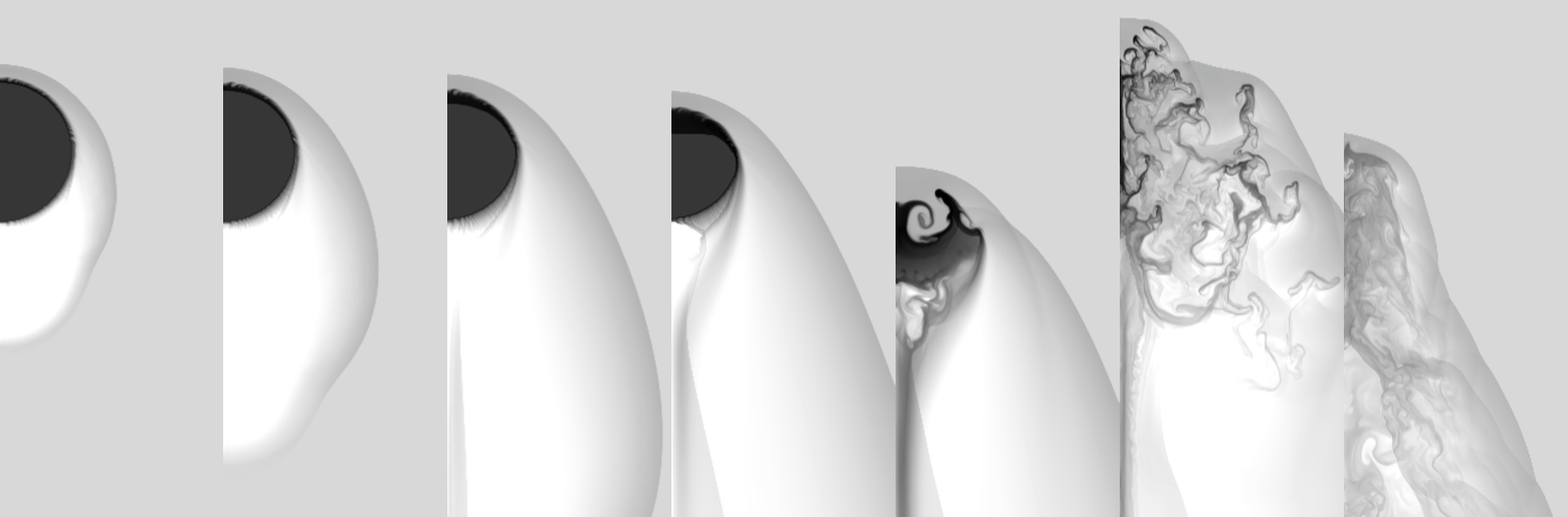} \\
\includegraphics[width=170mm]{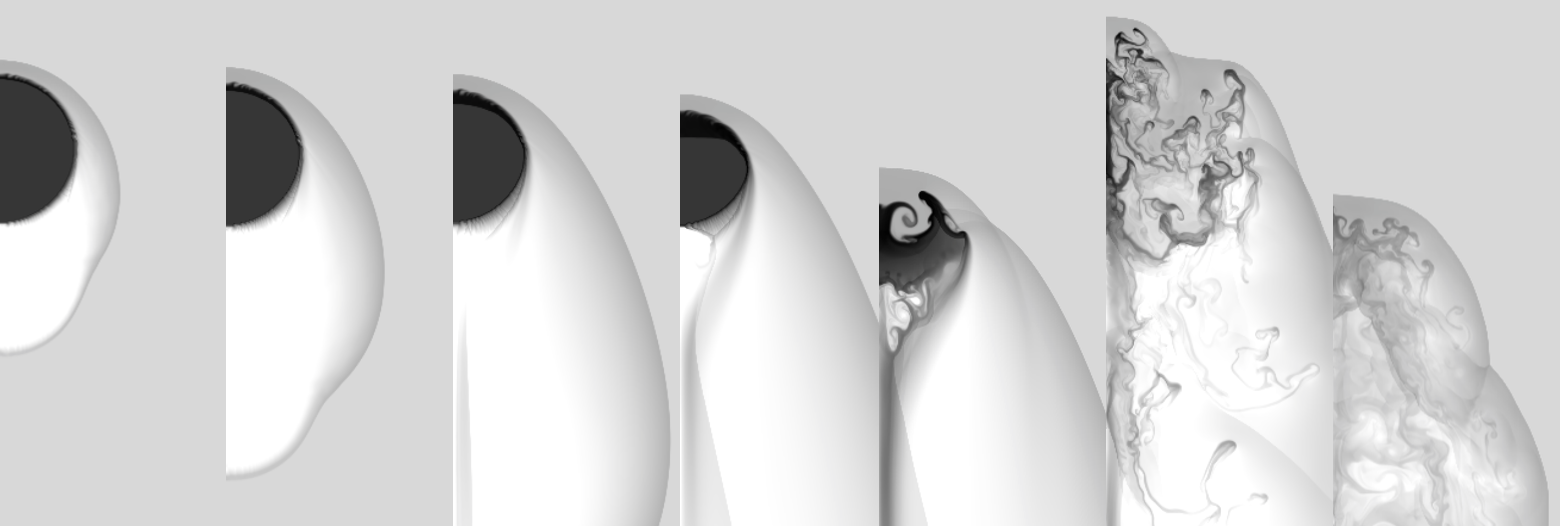}\\
\end{tabular}
\caption{The time evolution of the logarithmic density for models $c3wind1a$ (top row), $c3wind1b$ (middle row), and $c3wind1c$ (bottom row). The greyscale shows the logarithm of the mass density, scaled with respect to the ambient medium. The density scale used in this figure extends from 0 to 3.8. The evolution proceeds left to right with $t=0.07 \,t_{cc}$, $t=0.13 \,t_{cc}$, $t=0.25 \,t_{cc}$, $t=0.49 \,t_{cc}$, $t=1.84\,t_{cc}$, $t=3.10\,t_{cc}$, and $t=5.53 \,t_{cc}$. The first five frames in each set show the same region ($-5 < z < 2$, $0 < r < 3$, in units of $r_{c}$) so that the motion of the cloud is clear. The displayed region is shifted in the 6th frame of each set ($-13 < z < -1$, $0 < r < 5$) and the last frame ($-23 <  z < -11$, $0 < r < 5$) in order to follow the cloud.} 
\label{Fig3}
\end{figure*}

Interestingly, these high-$M_{wind}$ models initially form a thin, compressed, smooth tail of material ablated from the side and rear of the cloud (see panels 2, 3, and 4, corresponding to $t=0.13,0.25$ and $0.49\,t_{cc}$, in each set of Fig.~\ref{Fig3}), whereas, as already noted, the cloud in model $c3wind1$ forms instead a low-density turbulent wake. The cause of this is the way the flow moves around the cloud edge. In model $c3wind1$ the wind flows much closer to the cloud all the way around its edge. However, in model $c3wind1a$ the stronger bow shock deflects some of the flow away from the cloud edge, whilst the cavity serves to restrict the flow immediately behind the cloud. Thus, there is a slower removal of material from the cloud in the latter case. In addition, in model $c3wind1a$, the flow converges on the $r=0$ axis, which serves to focus cloud material at this point, whereas in model $c3wind1$ the flow changes direction and pushes upwards into the rear of the cloud. There is much less focusing of cloud material on the $r=0$ axis in this case and, thus, the tail of cloud material is much broader. This behaviour also differs from the comparable models in Paper I.

The fragments of cloud core in all higher velocity wind models remain encased in the strong bow shock. Furthermore, it is clear from Fig.~\ref{Fig3} that the cloud core in model $c3wind1c$ has travelled much further in the axial direction than that in model $c3wind1a$ (cf. the final panel in each set).

\subsection{Statistics}
\begin{figure*} 
\centering
\renewcommand{\arraystretch}{3}
  \begin{tabular}{cc}
   \resizebox{70mm}{!}{\includegraphics{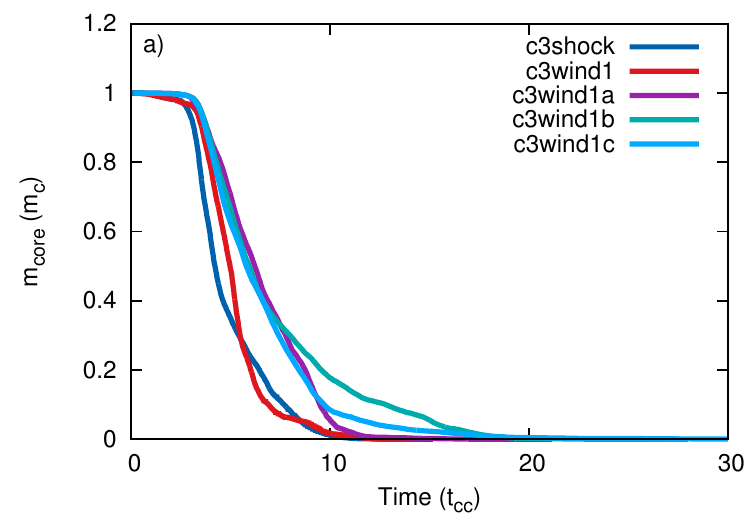}} &
     \resizebox{70mm}{!}{\includegraphics{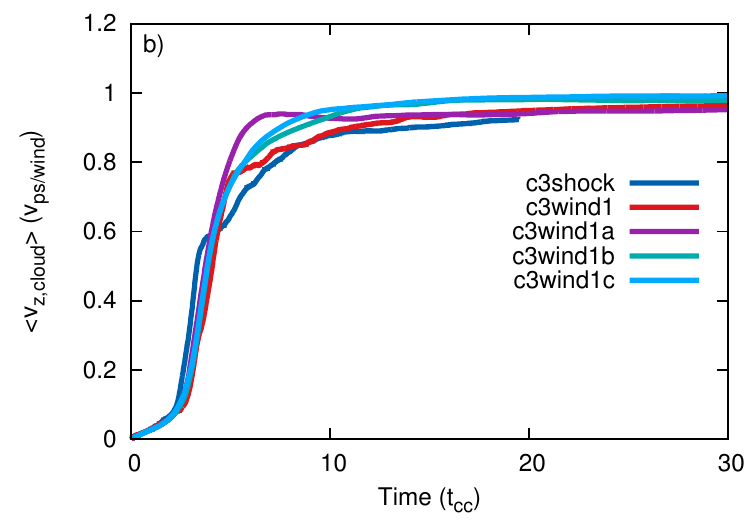}}   
     \multirow{2}{*}{     
    } \\
     \resizebox{70mm}{!}{\includegraphics{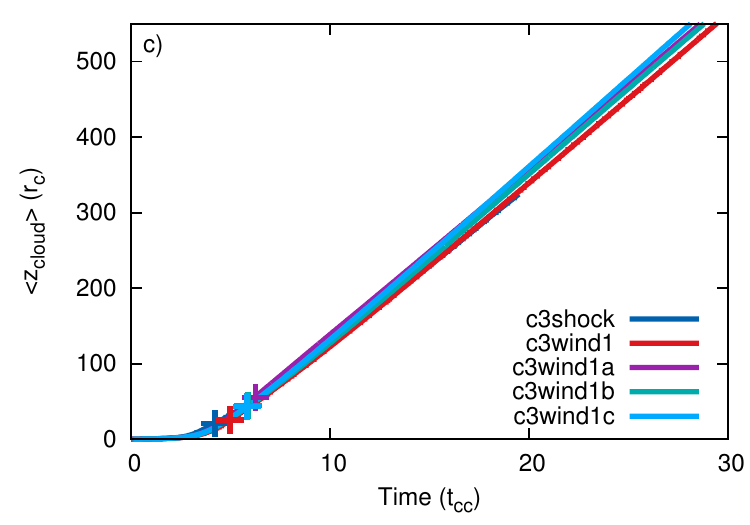}} &
     \resizebox{70mm}{!}{\includegraphics{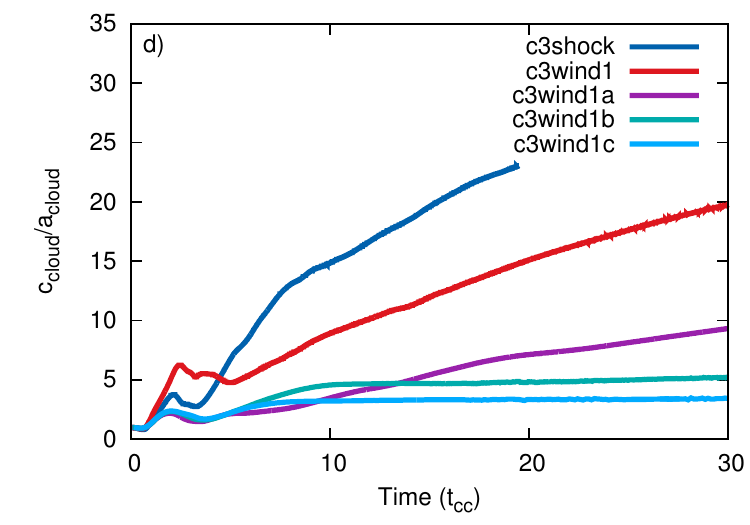}}   \\
      \resizebox{70mm}{!}{\includegraphics{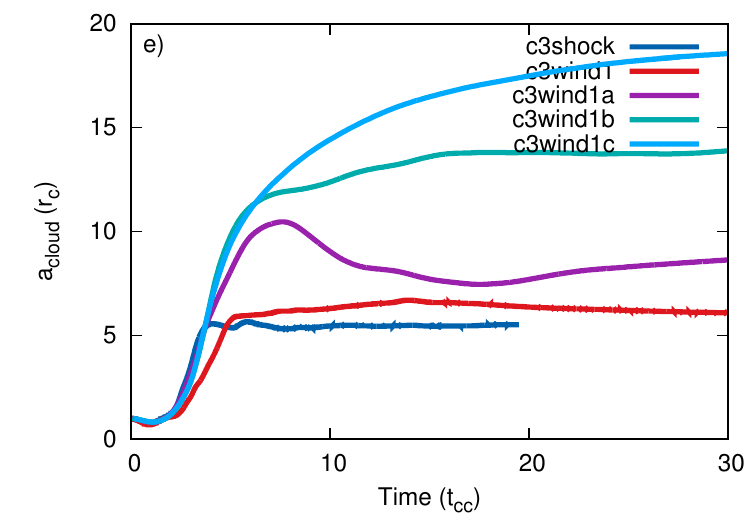}} &
     \resizebox{70mm}{!}{\includegraphics{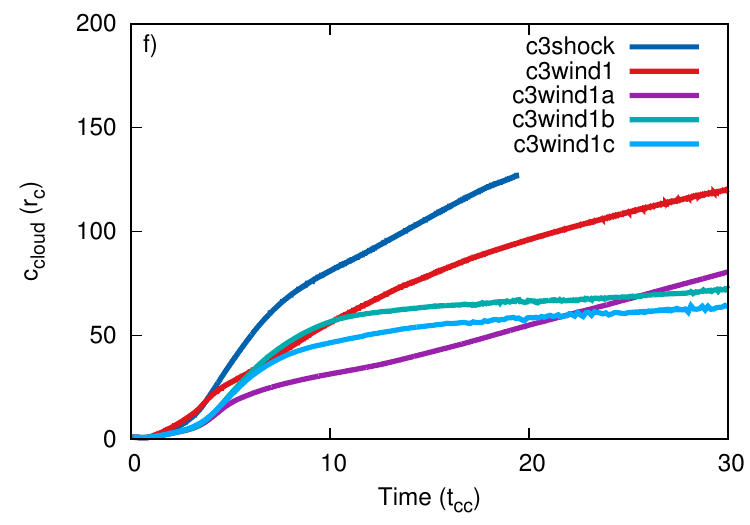}}   \\     
    \end{tabular}
  \caption{Time evolution of (a) the core mass of the cloud, $m_{core}$, (b) the mean velocity of the cloud in the $z$ direction, $\langle v_{z} \rangle$, (c) the centre of mass in the axial direction, $\langle z \rangle$, (d) the ratio of cloud shape in the axial and transverse directions, $c_{cloud}/a_{cloud}$, (e) the effective transverse radius of the cloud, $a_{cloud}$, and (f) the effective axial radius of the cloud $c_{cloud}$. Note that panel (c) shows the position of the centre of mass of each cloud at $t=t_{mix}$ (indicated by the respectively-coloured crosses). In addition, the behaviour of the cloud in model $c3shock$ after $t\approx 20\, t_{cc}$ has not been included in any of the above panels since the cloud material drops below the $\beta=2/\chi$ threshold at late times (see \S 2.2).}
  \label{Fig4}
  \end{figure*}

We now explore the evolution of various global quantities of the interaction for both the shock-cloud and wind-cloud models. Figure~\ref{Fig4} shows the time evolution of these key quantities, whilst Table~\ref{Table3} lists various time-scales taken from these simulations.

\begin{table}
\centering
 \caption{A summary of the cloud-crushing time, $t_{cc}$, and key time-scales, in units of $t_{cc}$, for the simulations investigated in this work. Note that the value for $t_{drag}$ given here is calculated using the definition given in \S2.3, whilst $t_{drag}'$ is the time when $\langle v_{z, cloud}\rangle = v_{ps}/e$, where $v_{ps}$ is the post-shock (or wind) speed in the frame of the unshocked cloud.}
 \label{Table3}
 \begin{tabular}{@{}lccccc}
  \hline
  Simulation & $t_{cc}$ & $t_{drag}$ & $t_{drag}'$ &$t_{mix}$ & $t_{life}$ \\
  \hline
c3shock  & 2.331 & 4.86	& 3.04  & 4.21 & 10.2 \\
c3wind1  & 2.331 & 4.46 &  3.69 & 4.97 & 10.9 \\
c3wind1a  & 0.737 & 4.16 &  3.40  & 6.23 & 11.7  \\
c3wind1b  & 0.233 & 4.25 &  3.43  & 5.87 & 17.8 \\
c3wind1c  & 0.074 & 4.38 & 3.53  & 5.82 & 17.6 \\
  \hline
 \end{tabular}
\end{table}

Figure~\ref{Fig4}(a) shows the time evolution of the core mass of the cloud in each of the simulations. It can be seen that models $c3shock$ and $c3wind1$ are closer in their behaviour than either of them is to the higher wind Mach number simulations (which, however, are more closely converged to each other as expected from Mach scaling considerations). The cloud core in model $c3shock$ drops to 50\% of its initial value more quickly than that of model $c3wind1$ due to the faster transverse expansion of the cloud in the former case. However, the greater lateral expansion of the cloud in model $c3wind1$ at later times, and hence its greater effective cross-section, means that it then loses mass from its core at a faster rate, between $t=5.5$ and $8.3\,t_{cc}$.

The rate of mass loss of model $c3shock$ is considerably faster than the comparable model $c1shock$ in Paper I where the cloud core survived until $t\approx 24\,t_{cc}$. In contrast, the mass loss is very similar between models $c3wind1$ and $c1wind1$, the cores of which are both destroyed by $t\approx 15\, t_{cc}$. In the shock-cloud cases, the turbulent wake evident in model $c3shock$ serves to hasten the rate of mass loss, compared to model $c1shock$ which lacked such a wake. The cloud core in model $c1wind1$ becomes compressed by secondary shocks which travel upwards from the rear of the core, and it develops filamentary structures at the rear much earlier than the cloud in model $c1shock$. Thus, the rate of core mass loss in $c1wind1$ is quicker than that in model $c1shock$, and comparable to $c3wind1$ where the core fragments.

The clouds in models $c3wind1a$, $c3wind1b$, and $c3wind1c$ are the slowest of the clouds in Fig.~\ref{Fig4}(a) to lose mass and have a slightly shallower mass-loss curve due to the lack of a turbulent wake prior to core fragmentation. These models have very similar core-mass profiles until $t\simeq 8\, t_{cc}$, when random fluctuations cause subsequent divergence in the evolution of $m_{core}$. The mass loss rate is considerably quicker for the wind-cloud models in the current paper than those in Paper I since the former fragment whilst the latter remain much more intact over a longer period before becoming mixed into the flow. Therefore, the cloud cores in the current paper have much steeper mass loss curves. 

The values of $t_{life}$ given in Table~\ref{Table3} are further confirmation that the cloud lifetime (normalised by $t_{cc}$) \textit{increases} with Mach number in wind-cloud interactions \citep{Scannapieco15, Goldsmith17}, as opposed to decreasing with Mach number in shock-cloud interactions \citep[e.g.][]{Pittard10, Pittard16b}, until Mach scaling kicks in at high Mach numbers, whereupon $t_{life}/t_{cc}$ approaches a constant value. Previous shock-cloud studies \citep[e.g.][]{Pittard16b} have shown that at low shock-Mach numbers dynamical instabilities on the cloud edge are slow to form; however, such instabilities are more prevalent as the Mach number increases, thus allowing the cloud to be shredded and mixed into the flow more rapidly, and reducing the cloud lifetime. However, in the wind-cloud case such instabilities are retarded as the wind Mach number increases, lessening the stripping of cloud material from the edge of the cloud in the higher $M_{wind}$ runs in Paper I and the current paper. Such dampening of the growth of KH instabilities and less effective stripping provide for a longer times-cale over which mass is lost.

The acceleration of the cloud is shown in Fig.~\ref{Fig4}(b). The cloud in model $c3wind1$ has a slightly slower acceleration than that in $c3shock$. Compared to Paper I, these two models show a slightly slower initial acceleration, due to the increased density of the cloud in these cases (for instance, the speed of the transmitted shock through the cloud is much slower). In addition, the non-smooth acceleration of both clouds between $t\approx 4-15\, t_{cc}$ acknowledges the change in shape of the cloud core away from the previous near-spherical morphology. The acceleration of the cloud in the higher $M_{wind}$ simulations initially follows that of the cloud in $c3wind1$. The acceleration of the cloud up to the asymptotic velocity is much smoother than seen in models $c3shock$ and $c3wind1$. The similar behaviour of the higher $M_{wind}$ simulations, as in Paper I, indicates the presence of Mach scaling. 

Figure~\ref{Fig4}(c) shows the time evolution of the cloud centre of mass in the axial direction. The movement of the centre of mass of the cloud in models $c3shock$ and $c3wind1$ is near identical. Models $c3wind1a$ to $c3wind1c$ differ very slightly in that the plot of the centre of mass of the cloud in these simulations is marginally steeper than that of the other two models from $t\approx 12\, t_{cc}$, indicating that they have moved downstream slightly further than the clouds in the other two models. Interestingly, this behaviour contrasts with that given in Paper I, where models $c3shock$ and $c1wind1$ had noticeably \textit{steeper} profiles compared to the higher $M_{wind}$ models.

\citet{Scannapieco15} found that clouds with $\chi \gtrsim100$ in a high-velocity flow were unable to be accelerated to the wind velocity before being disrupted, with clouds with a lower density contrast embedded in a high-velocity wind attaining much greater velocities. This suggests that clouds with high density contrasts would have difficulty in being moved across large distances before they are disrupted. We find that due to their large reservoir of mass, clouds with an initially high density contrast are able to significantly ``mass-load'' the flow, thus generating much longer-lived structures with density substantially greater than that of the background flow (see e.g. the last two time snapshots of each model in Fig.~\ref{Fig3}). These structures are able to move 100s of $r_{c}$ downstream from the original cloud position and acquire velocities comparable to the background flow speed. We find that this process is facilitated in high-velocity winds: the cloud in model $c3wind1c$ accelerates faster and is moved a greater distance than the cloud in model $c3wind1$. We note also that neither the complete mixing of cloud material, nor complete smoothing of the flow, are achieved in any of our simulations.

The time evolution of the shape of the cloud is presented in Fig.~\ref{Fig4}(d-f). In terms of the transverse radius of the cloud, $a_{cloud}$, the clouds in both $c3shock$ and $c3wind1$ show a modest expansion until $t\approx 4\,t_{cc}$ (not dissimilar to models $c1shock$ and $c1wind1$ in Paper I) before levelling out, coinciding with the moderate compression of the cloud in each case by the transmitted shock. The clouds in both models have a much greater expansion in the axial direction ($c_{cloud}$), coinciding with the formation of their turbulent wakes, in contrast to the behaviour found in Paper I where there was a much more modest axial expansion for the equivalent models \citep[cf. Fig.~\ref{Fig4}(f) with the same figure in][]{Goldsmith17}. In contrast, the cloud in $c3wind1c$ shows much less expansion in the axial direction (its axial radius nearly plateaus after $t\simeq 10\, t_{cc}$), whilst its expansion in the transverse direction is $3-4\times$ as large as the cloud in $c3shock$ and $c3wind1$. This is caused by the pressure and flow gradients resulting from the strong bow shock surrounding the cloud. Again, it can be seen that the cloud in model $c3wind1b$ behaves similarly to that in $c3wind1c$ in terms of the evolution of $c_{cloud}$, thus demonstrating Mach scaling.

\subsection{Time-scales}  
Table~\ref{Table3} provides normalised values for $t_{drag}$, $t_{mix}$, and $t_{life}$ for each of the simulations presented in this paper. Figure~\ref{Fig5} also shows the normalised values of $t_{drag}'$ and $t_{mix}$ as a function of the Mach number, and also in comparison to 2D inviscid shock-cloud simulations with $\chi=10^{3}$. The behaviour of each time-scale is now discussed in turn.

\subsubsection{$t_{drag}$}
First, we note that our wind-cloud simulations all have $t_{drag}/t_{cc} \approx 4.2-4.5$ (see Table~\ref{Table3}). These values are typically slightly greater than the values seen from the lower $\chi$ wind-cloud simulations in Paper~I, which spanned the range $3.3-4.3$. Thus, clouds with $\chi=10^{3}$ are accelerated by a wind slightly more slowly than those with $\chi=10$. This dependence is consistent with that also found in shock-cloud simulations \citep[see e.g.][]{Pittard10}, but in both cases the scaling is weaker than the $\chi^{1/2}$ scaling expected from a simple analytical model \citep[][]{Klein94,Pittard10}. We also find barely any Mach-number dependence to the values of $t_{drag}/t_{cc}$ in our wind-cloud simulations, when $\chi=10$ and $10^{3}$. This contrasts with the behaviour seen in shock-cloud simulations, where $t_{drag}/t_{cc}$ rises sharply at low Mach numbers \citep[e.g.][]{Pittard10,Pittard16b}.

\subsubsection{$t_{mix}$}
\begin{figure*} 
\centering
  \begin{tabular}{cc}
   \resizebox{70mm}{!}{\includegraphics{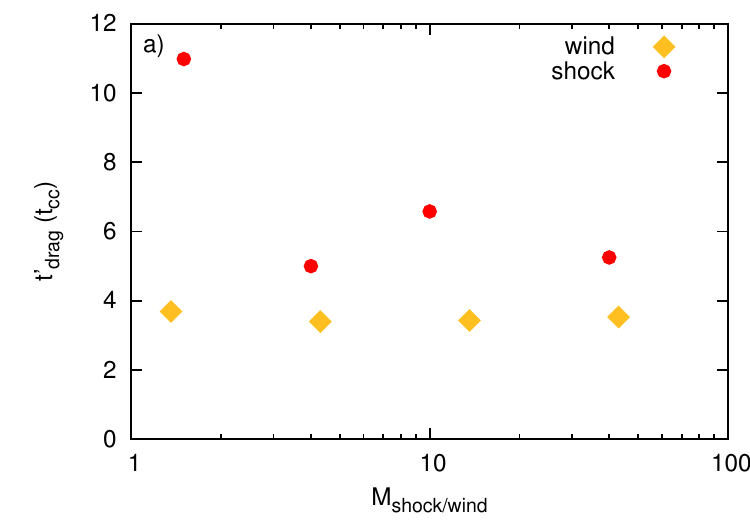}} &
     \resizebox{70mm}{!}{\includegraphics{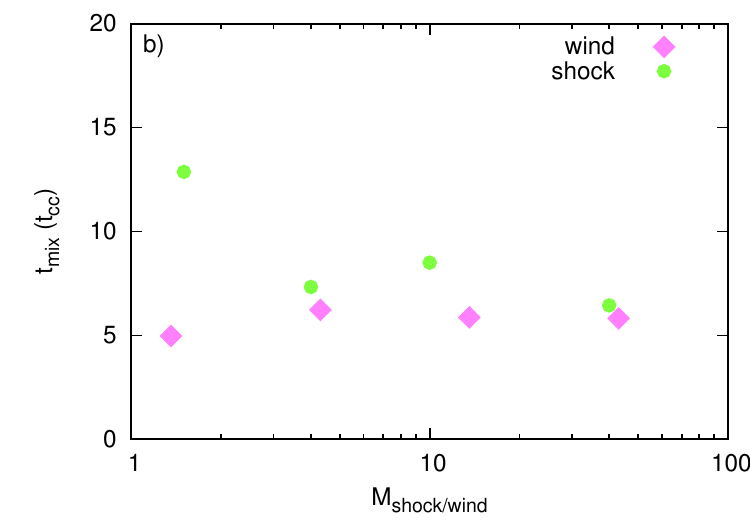}}   \\
    \end{tabular}
  \caption{(a) Cloud drag time, $t_{drag}'$, (gold diamonds) and (b) mixing time of the core, $t_{mix}$, (pink diamonds) as a function of the wind Mach number, $M_{wind}$ for the wind-cloud simulations. Also shown are the corresponding values from 2D inviscid simulations calculated for a shock-cloud interaction with $\chi=10^{3}$ ($t_{drag}$, red circles; $t_{mix}$, green circles). Note that in this figure, $t_{drag}'$ is defined as the time at which the mean cloud velocity, $\langle v_{z, cloud} \rangle= v_{ps}/e$, where $v_{ps}$ is the post-shock (or wind) speed in the frame of the unshocked cloud. This definition is consistent with \citet{Pittard10}, but differs from \citet{Klein94} and \citet{Pittard16b}. Thus, $t_{drag}' < t_{drag}$. See Table~\ref{Table3} for values of $t_{drag}$ calculated according to the definition given in \S2.3 of the current paper.}
   \label{Fig5}
  \end{figure*}

Table~\ref{Table3} and Fig.~\ref{Fig5} show that $t_{mix}/t_{cc}$ is almost independent of Mach number for the $\chi=10^{3}$ wind-cloud simulations presented in this paper. This behaviour contrasts with that from the $\chi=10$ wind-cloud simulations in Paper I, and the results of \citet{Scannapieco15}, where simulations with higher wind Mach numbers had significantly longer mixing times. Both behaviours contrast with the rapid rise in $t_{mix}/t_{cc}$ at low Mach numbers in shock-cloud simulations \citep{Pittard10,Pittard16b}! This clearly reveals very interesting diversity between these various interactions and motivates further studies of them. In particular, it is not clear why \citet{Scannapieco15} find longer mixing times with higher wind Mach numbers, when the current work does not, although there are a number of obvious avenues to investigate, including differences between the initial conditions and physics included, the effects of numerical resolution, and differences in the definition of mixing. As a final point, we note that Mach scaling is demonstrated in all of our work \citep{Pittard10,Pittard16b,Goldsmith17}, including the present.
 
Interestingly, Fig.~\ref{Fig5}(b) shows that the values of $t_{mix}/t_{cc}$ from the shock-cloud simulations (which \textit{do} show a Mach number dependence) appear to converge towards the Mach number-independent wind-cloud values as $M_{shock/wind}$ increases. This behaviour, although not quite so clear cut, may also be taking place for $t_{drag}'/t_{cc}$ too (see Fig.~\ref{Fig5}(a). Finally, we note that $t_{drag}'/t_{mix} \sim 0.6$ in our $\chi=10^{3}$ wind-cloud simulations (see Fig.~\ref{Fig5}).

\subsection{Comparison to existing literature}
As noted in Section~\ref{intro}, there is a lack of numerical studies in the literature that investigate the Mach-number dependence of wind-cloud interactions at high density contrast ($\chi \gtsimm 10^{3}$). Studies which consider high values of $\chi$ are often limited to a single value of $M_{wind}$ \citep[e.g.][]{Vieser07,Cooper09,Banda16}. Thus, it is difficult to draw any conclusions from the current literature as to the Mach-number dependence of $t_{mix}$ in wind-cloud simulations at high $\chi$. In fact, the only other wind-cloud study, to our knowledge, to investigate a range of Mach numbers at high $\chi$ is by \citet{Scannapieco15}. They find an increasing trend for $t_{mix}$ with $M_{wind}$, which is in disagreement with the results that we present here. This disagreement may be related to the different initial setup (their cloud is initially assumed to be in pressure equilibrium with the surrounding wind, whereas our cloud is under-pressured), or to the different physics employed (their simulation is radiative, whereas ours is adiabatic). In addition, there are numerical differences (e.g. 2D vs. 3D), and differences in the definition of mixing between their work and ours. Further investigation into the effect of these differences is needed.

In previous shock-cloud studies, \citet{Pittard10} and \citet{Pittard16b} showed that the ratio $t_{drag}'/t_{mix}$ was $\chi$-dependent\footnote{In these works, $t_{drag}$ is equivalent to $t_{drag}'$ in our current paper.}. To first order, the normalised mixing time-scale is independent of $\chi$, while the normalised drag time-scale increases weakly with $\chi$. Thus, clouds with low density contrasts are accelerated more quickly than they mix, while clouds with very high density contrasts tend to mix more efficiently than they are accelerated. At high Mach numbers ($M_{shock} \gtsimm 10$), \citet{Pittard16b} found that $t_{drag}'/t_{mix}$ increased from 0.14 when $\chi=10$, to 0.75 when $\chi=10^{3}$. Our current work now allows us to examine whether such behaviour is displayed in wind-cloud interactions. At high Mach numbers, Paper I showed that for $\chi=10$, $t_{drag}'/t_{mix} \approx 0.1$, while here we find $t_{drag}'/t_{mix} \approx 0.6$ for $\chi=10^{3}$. Thus, we find that mixing becomes relatively more efficient compared to acceleration for wind-cloud interactions as the cloud density contrast increases, in agreement with the behaviour seen in shock-cloud interactions.

\section{Summary and Conclusions}
This is the second part of a study comparing shock-cloud and wind-cloud interactions and the effect of increasing the wind Mach number on the evolution of the cloud. Our first paper \citep{Goldsmith17} investigated the morphological differences between clouds of density contrast $\chi=10$ struck by a shock and those embedded in a wind. Significant differences were found, not only between the morphology of the clouds themselves but also in terms of the behaviour of the external medium in each case. It was also the first paper to identify Mach scaling in a wind-cloud simulation and additionally found that clouds embedded in high Mach number winds survived for longer and travelled larger distances.

In this second paper, we have continued our investigation of shock-cloud and wind-cloud interactions, but this time have focussed on clouds with a density contrast of $\chi=10^{3}$. As in Paper I, we began our investigation by comparing wind-cloud simulations against a reference shock-cloud simulation with a shock Mach number $M=10$ ($c3shock$). Our standard wind-cloud simulation ($c3wind1$) used exactly the same cloud embedded in the same flow conditions. On comparing the two simulations, we find only minor morphological differences between the clouds in each simulation whilst the transmitted shock progresses through the cloud. After the transmitted shock has exited the cloud, we find that the cloud in both models begins to develop a low-density turbulent wake. The evolution of the two clouds begins to diverge after this time, and the morphology and properties of the cloud become increasingly different with time. For instance, the development of the wake differs significantly between the two models: the cloud core in model $c3shock$ does not fragment but is drawn out along the $r=0$ axis, whilst that in model $c3wind1$ does fragment and eventually disrupts the evolution of the wake.

On increasing the wind Mach number, we find that a supersonically-expanding cavity quickly forms at the rear of the cloud, similar to the higher $M_{wind}$ simulations in Paper I. This is followed by a smooth, compressed, thin, but short-lived tail of cloud material which forms behind the cloud. This narrow tail arises from the focusing of the flow around and behind the cloud. Neither the cavity, nor the subsequent narrow tail, are seen in models $c3shock$ and $c3wind1$, or the comparable models in Paper I at lower $\chi$. In all of our new wind-cloud simulations, the cloud eventually fragments and mass-loads the flow.  

In Paper I, we demonstrated the presence of Mach scaling in wind-cloud simulations for the first time. Our new results shown here provide further evidence of this effect. For example, the clouds in the higher Mach number simulations are all morphologically very similar (cf. each set of panels in Fig.~\ref{Fig3}), and evolve closely until ``random'' perturbations caused by the different non-linear development of instabilities from numerical rounding differences in the simulations eventually cause them to diverge.

We also find that clouds with density contrasts $\chi > 100$ can be accelerated up to the velocity of the wind and travel large distances before being disrupted, in contrast to the findings of \citet{Scannapieco15}. For instance, in model {\it c3wind1a}, the cloud reaches 90\% of $v_{wind}$ by $t = t_{mix}$, at which time it has moved downstream $\approx 50\,r_{c}$. However, the flow remains structured and complete mixing is not achieved.

Our work has helped to reveal a rich variety of behaviours depending on the nature of the interaction (shock-cloud or wind-cloud) and the cloud density contrast. In shock-cloud interactions, both the normalised cloud mixing and drag times increase at lower Mach numbers, but are independent of Mach number at higher Mach numbers - i.e. they show Mach scaling \citep[see][]{Klein94,Pittard10,Pittard16b}. The drag time also increases weakly with $\chi$, but $t_{mix}/t_{cc}$ does not. In contrast, wind-cloud interactions with $\chi=10$ show an almost Mach-number-independent drag time, but a strong rise in $t_{mix}/t_{cc}$ with Mach number until $M_{wind} \sim 20$, whereupon $t_{mix}/t_{cc}$ plateaus as Mach-scaling is reached \citep{Goldsmith17}. Our current work reveals another type of behaviour: wind-cloud interactions with $\chi=10^{3}$ show almost Mach-number-independent drag and mixing times. Comparison of the current work with \citet{Goldsmith17} also reveals that the normalised cloud mixing time at high Mach numbers is shorter at higher values of $\chi$ in our wind-cloud simulations, which is opposite to the $\chi$-dependence seen in shock-cloud interactions where $t_{mix}/t_{cc}$ is essentially independent of $\chi$, and at most very weakly increases with it \citep{Pittard10,Pittard16b}. Finally, we find that the Mach number dependent values of $t_{drag}'$ and $t_{mix}$ for shock-cloud simulations at $\chi=10^{3}$ converge towards the Mach-number-independent time-scales of comparable wind-cloud simulations.

That shock-cloud and wind-cloud interactions display such richness of behaviour demands further investigation. In particular, there is a need to address some of the discrepancies which currently exist between different studies.

\section*{Acknowledgements}
This work was supported by the Science \& Technology Facilities Council [Research Grants ST/L000628/1 and ST/M503599/1]. We thank S. Falle for the use of the \textsc{mg} hydrodynamics code used to calculate the simulations in this work. The calculations used in this paper were performed on the DiRAC Facility which is jointly funded by STFC, the Large Facilities Capital Fund of BIS, and the University of Leeds. The data associated with this paper are openly available from the University of Leeds data repository. \url{https://doi.org/10.5518/221}

\end{document}